# DYNAMIC APERTURE OPTIMIZATION FOR THE DAΦNE UPGRADE

Eugene Levichev, Pavel Piminov [1], Pantaleo Raimondi, Mikhail Zobov [2]

[1] BINP SB RAS, Lavrentiev prospect 11, 630090 Novosibirsk, Russia
[2] INFN-Laboratori Nazionali di Frascati Via E. Fermi 40, I-00044 Frascati, Italy

## Abstract

Recently proposed novel idea of "crabbed waist" beam-beam collisions will be tested at DAΦNE during the collider run for the Siddharta experiment. In order to achieve the goal luminosity, large dynamic aperture is a matter of primary importance. A new method of a dynamic aperture optimization based on step-by-step chromaticity compensation with choosing the "best" sextupole pair at each step was applied to the DAΦNE upgrade lattice. Several tune points were considered taking into account both high luminosity and large dynamic aperture. Algorithm and results of optimization will be presented.

# 1 INTRODUCTION

Recently a novel idea of a "crabbed waist" beam-beam collision at large crossing angle was suggested as a way for luminosity increasing by one or two orders of magnitude [1]. Later this idea was proposed for the DAΦNE collider upgrade to test the "crabbed waist" concept and to enhance the luminosity of colliding beams for the Siddharta experiment [2].

In order to achieve the goal luminosity, a dynamic aperture of the machine should be sufficiently large, otherwise strong beam-beam effects, which cause increase of particles population in the beam tails, will lead to a reduction of beam lifetime and luminosity degradation. The beam-beam simulation for the upgraded DAΦNE has shown that the size of the dynamic aperture required to obtain high luminosity should be larger than $15\sigma_x$ in the horizontal plane and $150\sigma_y$ in the vertical one [5]. Requirement for the momentum acceptance is $A_{\Delta E/E} \geq 0.5\%$ .

To compensate the natural chromaticity and at the same time to optimize a dynamic aperture of a storage ring, two possible approaches may be considered. The first one uses theoretical tools to estimate and to reduce strength of nonlinear perturbation (resonance driving terms, action invariant smear, nonlinear detuning coefficients, etc.). The following problems complicate practical use of this approach: (a) there is no a single estimate for nonlinear perturbation valid for all cases and for all betatron tunes; (b) there is no direct relation between perturbation strength and the size of dynamic aperture.

The second approach does not use any theoretical models; instead of that it is based on general methods of numerical optimization. In the following we apply such algorithm choosing "the best" pairs of sextupole magnets for the chromaticity compensation to the upgraded DAΦNE lattice. The algorithm is simple and effective, does not require excessive running time and can be applied for an arbitrary lattice.

# 2 ALGORITHM

We propose to correct the chromaticity by $N$ small steps along the vector $\xi = (\xi_{x0}, \xi_{y0})$ as it is shown in Fig.1. At each step $1/N$-th fraction of the horizontal and vertical chromaticity is compensated by a single (in some sense the best for this particular step) pair of focusing and defocusing sextupoles $(SF_i, SD_j)$ .

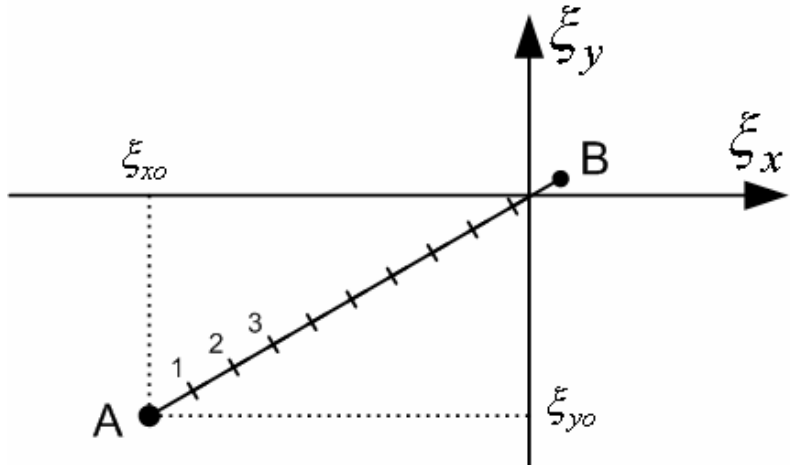

Fig.1 Step-by-step chromaticity compensation. A and B indicate initial and final points respectively.

To find the best pair of sextupoles, we try all possible $(SF, SD)$ - combinations and the pair demonstrating the largest dynamic aperture is fixed at this step. If $N_{SF}$ and $N_{SD}$ are the number of focusing and defocusing sextupoles, then $N_{SF} \times N_{SD}$ combinations have to be looked through at every step.

At the next steps the procedure is repeated until the chromaticity will reach the desired value.

As the dynamic aperture represents particle stable motion area with complicated and rather ambiguously determined boundary, an important problem is fast and reliable comparison of different apertures, provided by sextupole pairs tested at the particular step. Several functional criteria have been studied: the DA area, the area of ellipse inscribed into the DA boundary, the DA area normalized by the length of the boundary curve, etc. Weight factors can be introduced if there are some particular goals: for instance, increasing of the horizontal aperture while keeping the vertical aperture equal to the mechanical one (say, limited by small-gap undulator). Actually, it is difficult to indicate the only criterion because its effectiveness is usually defined by a specific task.

Once the chromaticity is corrected, we optimize DA further exploiting sextupoles placed in the dispersion-free sections. A gradient search algorithm is used for this purpose.

The algorithm may be naturally extended for increasing the off-momentum aperture: instead of a single DA with $\Delta p / p = 0$ several DAs with specified $(\Delta p / p)_i$ are optimized and no modifications are required.

## 3    OPTIMIZATION RESULTS

We have started with the DAΦNE lattice [3], which main parameters are listed in Table 1 and optical functions are plotted in Fig.2. Later on we shall use this lattice as a reference one and denote it as `DAΦNE_Siddharta_2007_0`.

Table 1: Main parameters of `DAΦNE_Siddharta_2007_0`.

| Betatron tunes | $Q_x/Q_y$ | 5.103/5.179 |
|---|---|---|
| Compaction factor | $\alpha$ | 0.0193 |
| Damping times (ms) | $\tau_x / \tau_y / \tau_s$ | 39.8/34.5/16.1 |
| Horizontal emittance (nm-rad) | $\varepsilon_x$ | 390 |
| Energy spread | $\sigma_E/E$ | $3.9 \times 10^{-4}$ |
| Natural chromaticity | $\xi_x / \xi_y$ | -3.2/-25.9[*] |
| IP1 betas (m) | $\beta^*_x / \beta^*_y$ | 0.2/0.006 |
| Beam size at IP1 (μm) | $\sigma^*_x / \sigma^*_y$ | 284/3.4 |

[*] All sextupoles are switched off except for the "crab waist" ones (placed in zero dispersion straights) and strong sextupole terms produced by shaped iron cap in the terminal poles of damping wigglers [4].

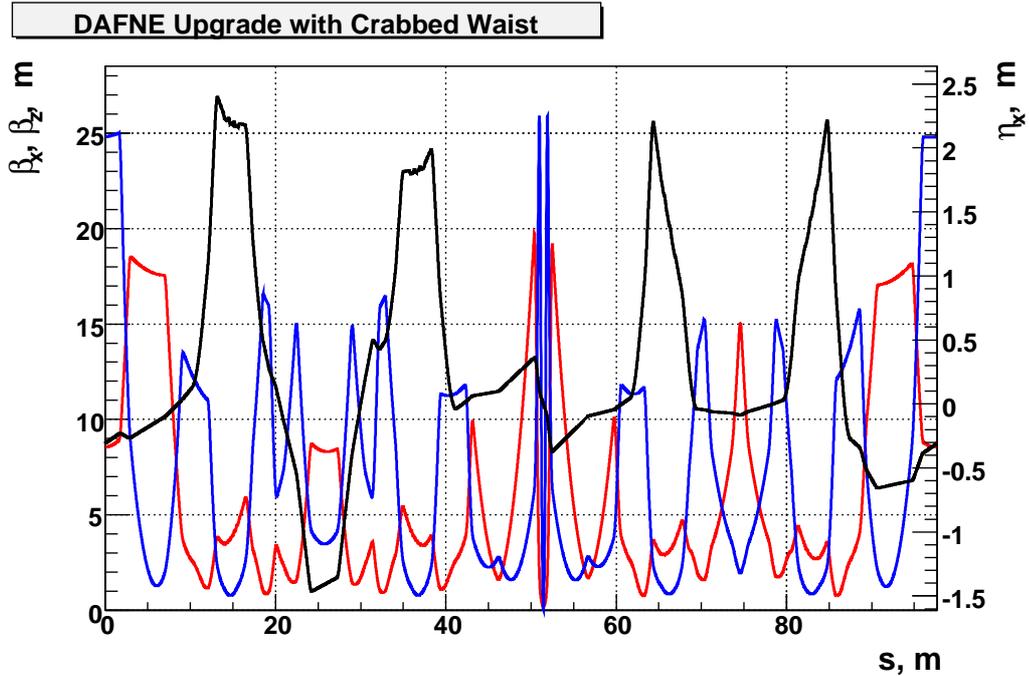

Fig.2 Lattice functions of `DAΦNE_Siddharta_2007_0`.

Nonlinear lattice elements include:

- two strong "crab waist" sextupole magnets *SXPPS101* and *SXPPL104* ($L = 0.5$ m, $B'' / B\rho = 136$ m$^{-3}$),

- set of chromatic sextupoles compensating the natural chromaticity to $\xi_x = -1$, $\xi_y = -2$,

- three vertically focusing sextupoles *SXPPS201*, *SXPPL201* and *SXPPS204* , which are located in the (almost) dispersion-free straight sections and could be considered as harmonic sextupoles,

- nonlinear components in 4 damping wigglers according to [4].

Details of the sextupole magnet parameters are given in Table 2.

Table 2: **DAΦNE_Siddharta_2007** sextupole magnet parameters

The legend:

**DS0** = **DAΦNE_Siddharta_2007_0**,{5.103, 5.179}/non optimized
**DS0/Opt** = **DAΦNE_Siddharta_2007_0**,{5.103, 5.179}/optimized
**DS1** = **DAΦNE_Siddharta_2007_1**,{5.105, 5.160}/optimized
**DS2** = **DAΦNE_Siddharta_2007_2**,{5.131, 5.116}/optimized

| Name | L(cm) | B"/BR (m$^{-3}$) | | | |
|------|-------|------|---------|------|------|
| | | DS0 | DS0/Opt | DS1 | DS2 |
| SXPPS101 | 10,00 | 136,11 | 136,11 | 136,11 | 136,11 |
| SXPPS102 | 15,00 | -60,85 | -29,10 | -95,78 | -40,84 |
| SXPPS103 | 15,00 | 40,78 | 8.29 | 40,51 | 15,46 |
| SXPPS201 | 10,00 | 0,00 | -18.95 | -19,50 | 0,38 |
| SXPPS202 | 15,00 | 40,78 | 45.84 | 43,30 | 30,24 |
| SXPPS203 | 15,00 | -47,47 | -75.55 | -45,76 | -95,97 |
| SXPPS204 | 10,00 | 0,00 | 6,25 | 6,83 | 1,17 |
| SXPPL201 | 10,00 | 0,00 | -13.42 | -3,06 | 0,71 |
| SXPPL202 | 15,00 | -47,47 | -38.04 | -15,90 | -52,10 |
| SXPPL203 | 15,00 | 10,68 | 9.68 | 3,67 | 10,48 |
| SXPPL204 | 10,00 | 23,87 | 12.02 | 8,85 | 12,28 |
| SXPPL100 | 10,00 | -8,38 | -11.9 | 0,80 | -33,12 |
| SXPPL101 | 10,00 | 23,87 | 8.69 | 3,87 | 1,68 |
| SXPPL102 | 15,00 | 10,68 | 22,39 | 21,85 | 10,85 |
| SXPPL103 | 15,00 | -60,85 | -90,89 | -78,32 | -48,96 |
| SXPPL104 | 10,00 | -136,11 | -136,11 | -136,11 | -136,11 |

Fig.3 shows the dynamic aperture of the **DAΦNE_Siddharta_2007_0** with the chromaticity compensated to $\xi_x = -1, \xi_y = -2$, while Fig.4 presents the horizontal phase space portrait, which is typical for the case when two resonances $\nu_x = n$ and $3\nu_x = 3n$ take place.

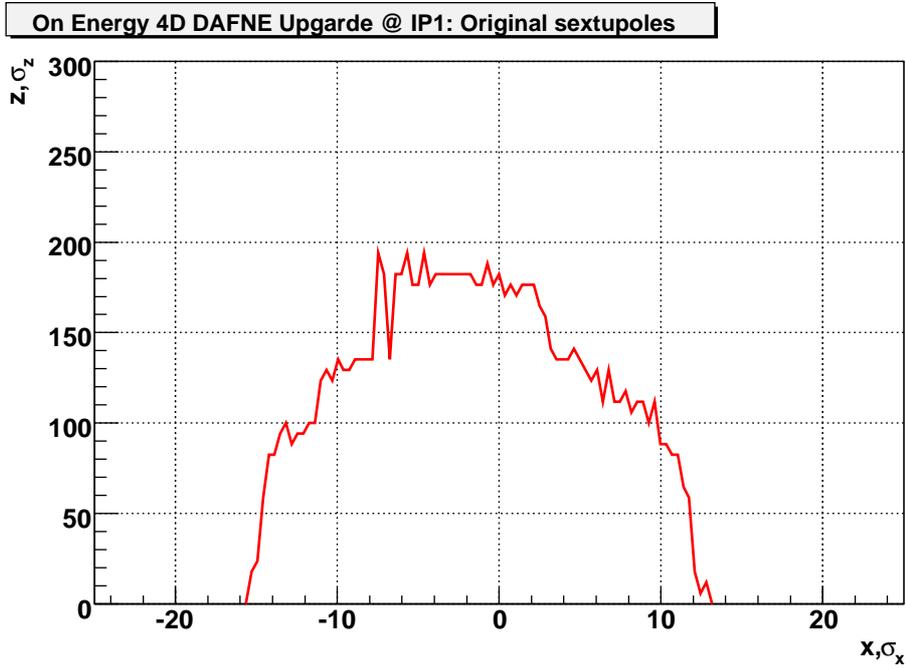

Fig.3: Dynamic aperture of **DAΦNE_siddharta_2007_0.**

All plots are performed for the IP1 azimuth and at this point the dynamic aperture in terms of sigma is equal to $A_x \approx {}^{+13\sigma_x}_{-16\sigma_x}$ and $A_y \approx 180\sigma_y$.

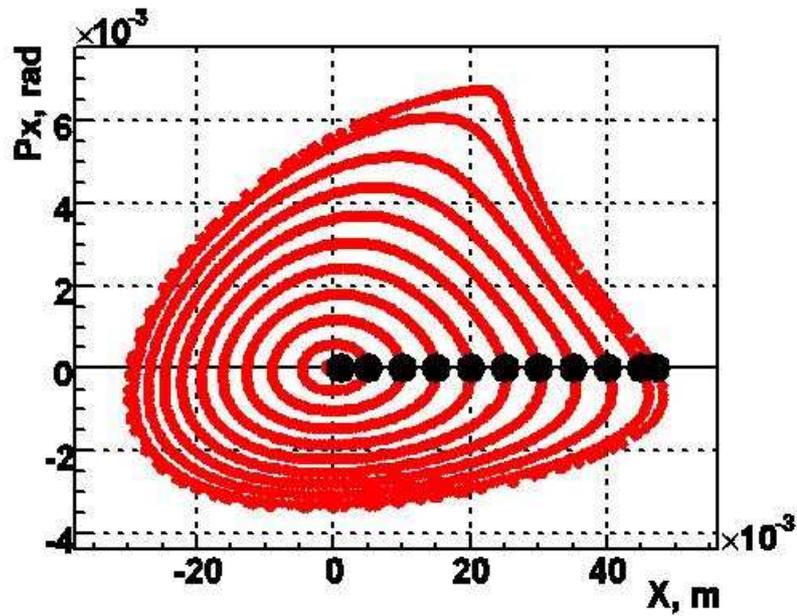

Fig.4: Horizontal phase curves of **DAΦNE_siddharta_2007_0.**

The DAΦNE dynamic aperture optimization has been performed according to the following scenario:

- The DA optimization at the original tune point {5.103, 5.179} by the "best pair" method.

- The DA tune scan in the vicinity of the original tune point in order to look for a larger aperture. At this point we have to superpose good DA region with high luminosity region according to the luminosity scan [2].

- Re-optimization of the DA at the new tune point(s).

- Investigation of the DA optimization with such options like octupole magnets energizing, modification of the wiggler nonlinear terms, etc.

### 3.1 DA optimization at the original tune point

30 $(SF, SD)$ pairs might be combined from the DAΦNE sextupole magnets and their optimization takes 0.5-2 hours on 2 GHz PC dependently on the internal optimization parameters.

At the original tune point {5.103, 5.179} the best pair algorithm yields the DA given in Fig. 5 and the sextupoles strength listed in the second column of Table 2.

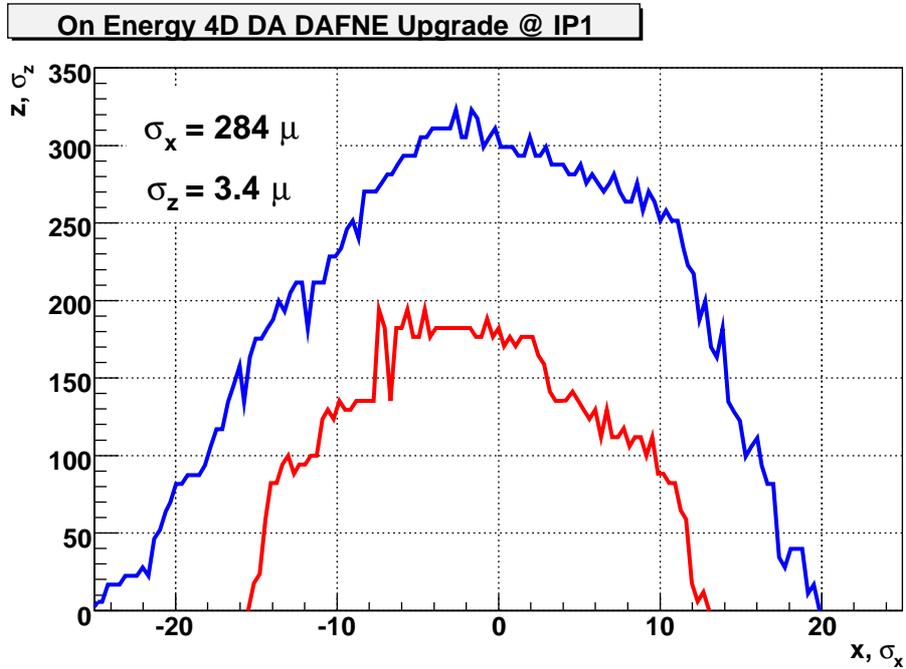

Fig.5 Optimized DA (blue) at the original tune point of **DAΦNE_Siddharta_2007_0**. The reference aperture is given in red.

At the IP1 the dynamic aperture now is equal to $A_x \approx ^{+16\sigma x}_{-23\sigma x}$ and $A_y \approx 250\sigma_y$.

*3.2 DA tune scan*

In order to adjust the betatron tunes for higher luminosity and at the same time for larger dynamic aperture, the luminosity tune scan (Fig.7) was compared with the DA tune scan (Fig.8). To smooth noisy and irregular shape of DA border line, we define the size of a stable motion area by semi axes of the ellipse inscribed into the DA contour as it is shown schematically in Fig.6, and just this definition was used for the plot in Fig.8.

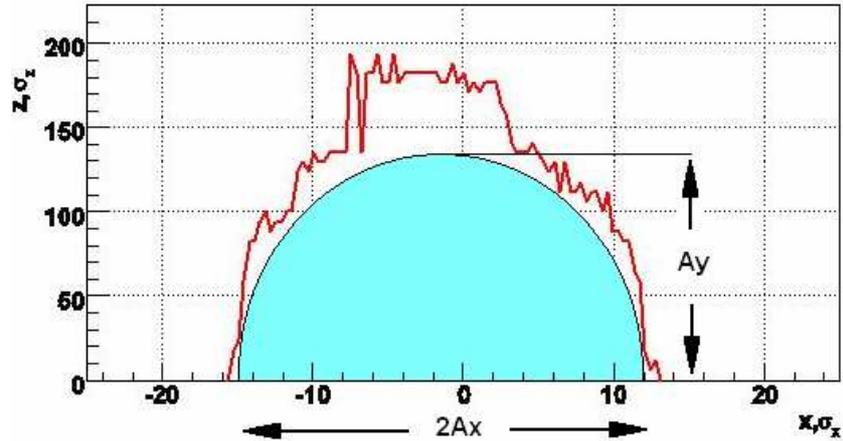

Fig.6 DA size definition (schematically).

The white area in the DA plot (Fig.8) corresponds to optically unstable solution because of particular choice of the QF and QD magnets to scan the betatron tunes. However, the scanned area seems quite enough to establish correlation between the DA and the luminosity.

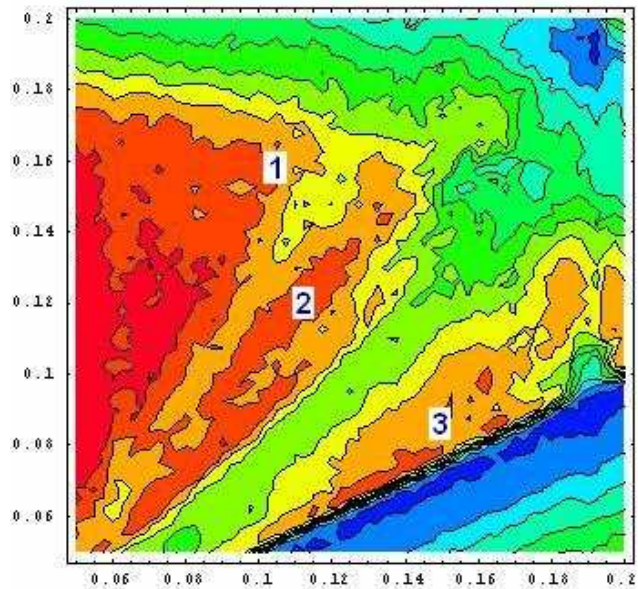

Fig.7 The luminosity tune scan.

Both scans clearly demonstrate the resonant lines structure reducing both the luminosity and the dynamic aperture. Among such resonances the strongest are $\nu_x - \nu_y$ (or $2\nu_x - 2\nu_y$ for the sextupole perturbation) and $\nu_x - 2\nu_y$. Although the difference resonance is intrinsically stable, strong coupling of two oscillation modes and large modulation of the betatron amplitudes may cause a reduction of dynamic aperture.

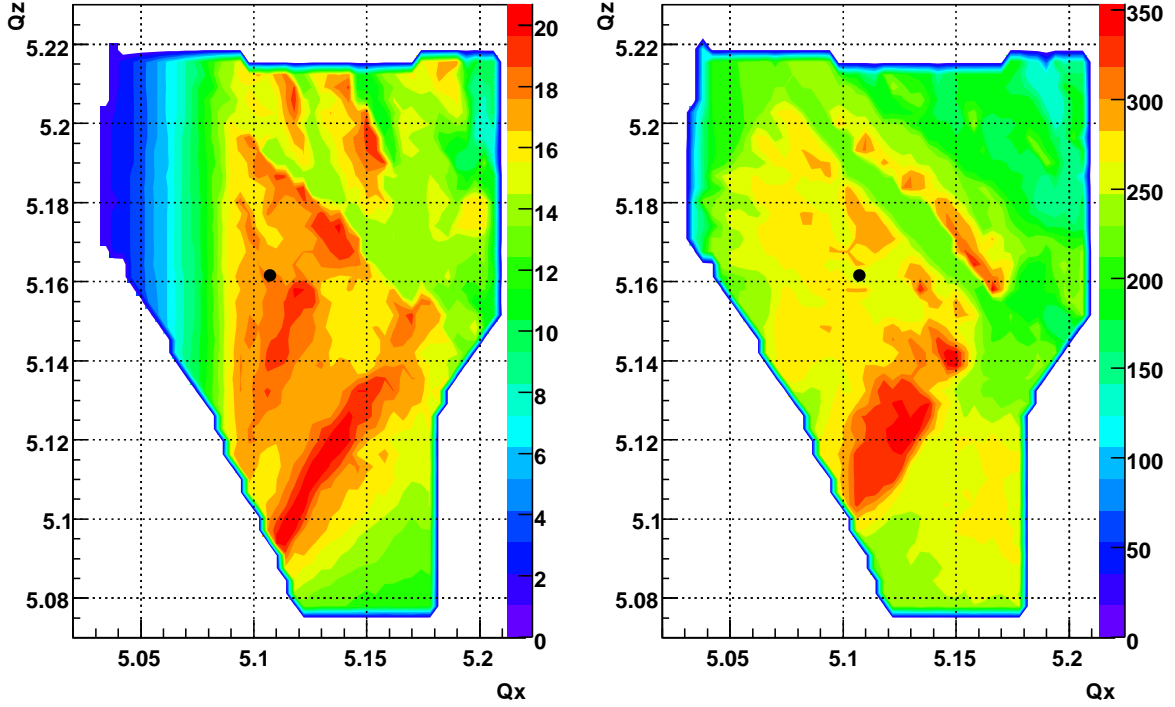

Fig.8 The dynamic aperture tune scan. Color indicates the DA size in term of sigma.

The scan in Fig.7 demonstrates 3 regions with maximum luminosity:

- Region 1 corresponds to high luminosity and sufficiently large dynamic aperture.
- Region 2 also provides large luminosity and dynamic aperture but this region is placed close to the main coupling resonance $\nu_x - \nu_y$ and reaching good parameters here for the real machine may be a matter of essential difficulty.
- Region 3 shows narrow luminosity ridge near the resonance $\nu_x - 2\nu_y$ but the dynamic aperture here is small.

### 3.3 DA re-optimization at the new tune points

From above consideration we have chosen two alternative tune points:

- {5.105, 5.160} from region 1 (**DAΦNE_Siddharta_2007_1**) and
- {5.131, 5.116} from region 2 (**DAΦNE_Siddharta_2007_2**)

to check and re-optimize the dynamic apertures. The results are shown in Fig.9. Both new tune points demonstrate the dynamic aperture significantly larger than the initial one (**DAΦNE_siddharta_2007_0**) and larger than that optimized (Fig.5). Re-optimized sextupole setting for each tune point is shown in Table 2. The summary of the dynamic aperture sizes before and after the optimization is given in Table 3.

Table 3. DAΦNE DA optimization summary

| Name | Tune Point | $N\sigma_x$ | $N\sigma_y$ | Comments |
|---|---|---|---|---|
| **DAΦNE_siddharta_2007_0** | 5.103, 5.179 | +13/-16 | 180 | Original |
| **DAΦNE_siddharta_2007_0** | 5.103, 5.179 | +16/-23 | 250 | Original, optimized |
| **DAΦNE_siddharta_2007_1** | 5.105, 5.160 | +20/-26 | 300 | |
| **DAΦNE_siddharta_2007_2** | 5.131, 5.116 | +20/-23 | 270 | Coupling resonance |

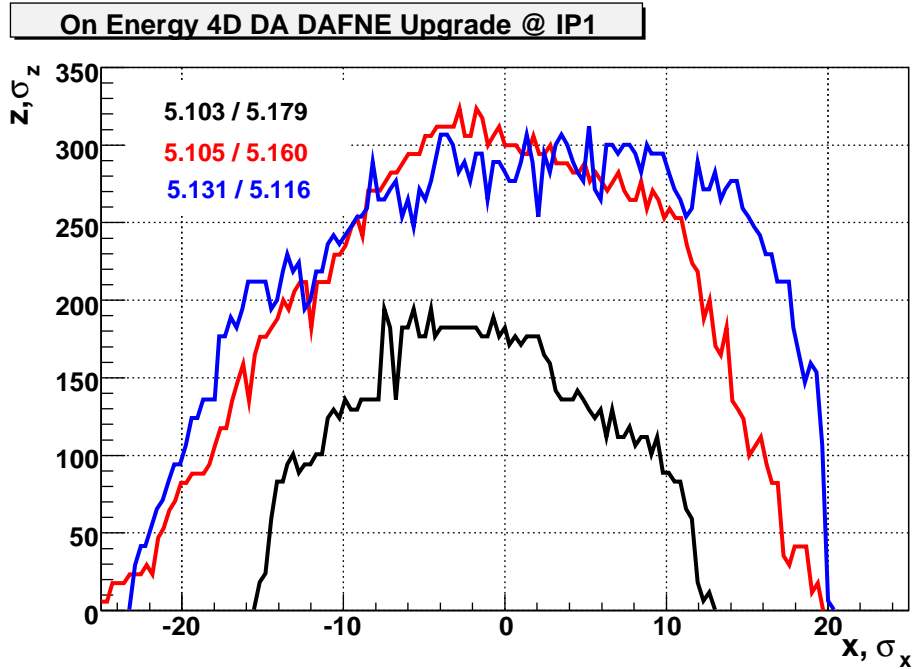

Fig.9. The optimized DA for different working points.

The best pair algorithm allows to optimize an off-momentum dynamic aperture but for our case it is not necessary because it seems to be large enough without any additional efforts (Fig.10): even for $\Delta p / p = 1\%$ the transverse DA is $\pm 5\sigma_x$ and $100\sigma_y$.

### 3.4 Other perturbation sources adjustment

Besides the regular sextupole magnets, DAΦNE contains other sources of nonlinear magnetic fields: three octupole magnets and damping wigglers. The damping wigglers have inner pole nonlinearities, which can hardly be modified and strong sextupole term in one of the terminal pole.

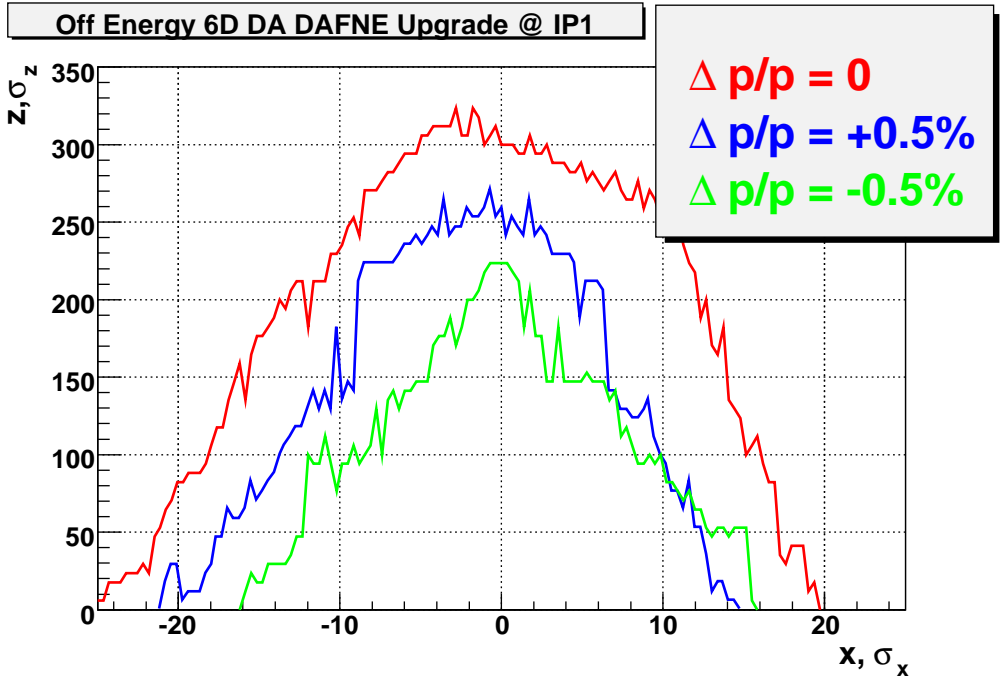

Fig.10. Off-momentum DA for **DAΦNE_siddharta_2007_1.**

This strong sextupole term introduced to correct the natural chromaticity is produced by superimposed iron plate whose shape, in principle, can be changed.

We have tried to optimize the sextupole arrangement for different set of octupole magnets but their influence on the dynamic aperture and on the final values of optimized sextupole magnets is negligible.

As for the strong sextupole component in the wiggler end pole, we have included it as a free parameter in the optimization process but its final value required to maximize the dynamic aperture turned out to be rather close to the original value (less than 10-15%) so it seems there is no need to modify it.

## 4    CONCLUSIONS

For the project of DAΦNE upgrade for the Siddharta run, arrangement of the sextupole magnets was optimized and the tune point was chosen from a viewpoint of high luminosity and large dynamic aperture. The "best pair" optimization method provides the dynamic aperture ≥20$\sigma$ in the horizontal direction and >250$\sigma$ in vertical one with the energy acceptance ~1%. These values seem quite satisfactory to provide high luminosity and successful experimental run. It is worth to note that one of the promising tune points {5.105, 5.160} practically coincides with that of the present DAΦNE run.